\documentstyle[12pt,epsf,epsfig,wrapfig]{article}

\newcommand{\bce}{\begin{center}}
\newcommand{\ece}{\end{center}}
\newcommand{\beq}{\begin{equation}}
\newcommand{\eeq}{\end{equation}}
\newcommand{\bea}{\begin{eqnarray}}

\newcommand{\eea}{\end{eqnarray}}

\newcommand{\ba}{\begin{array}}
\newcommand{\ea}{\end{array}}
\newcommand{\bDelta}{\mbox{\boldmath $\Delta$}}

\newcommand{\bkappa}{\mbox{\boldmath $\kappa$}}

\newcommand{\bfe}{{\bf e}}
\newcommand{\bb}{{\bf b}}
\newcommand{\br}{{\bf r}}

\newcommand{\bk}{{\bf k}}

\newcommand{\bn}{{\bf n}}

\newcommand{\bs}{{\bf s}}

\newcommand{\bV}{{\bf V}}

\def\lsim{\mathrel{\rlap{\lower4pt\hbox{\hskip1pt$\sim$}}
    \raise1pt\hbox{$<$}}}         %greater than or approx. symbol
\def\gsim{\mathrel{\rlap{\lower4pt\hbox{\hskip1pt$\sim$}}
    \raise1pt\hbox{$>$}}}         %greater than or approx. symbol
\def\beq{\begin{equation}}
\def\eeq{\end{equation}}
\def\bea{\begin{eqnarray}}
\def\eea{\end{eqnarray}}

\begin{document}

\begin{center}

{\bfseries IMPACT OF SATURATION ON S-CHANNEL HELICITY
NONCONSERVATION FOR DIFFRACTIVE VECTOR MESONS }

\vskip 5mm

I.P. Ivanov$^{1}$, \underline{N.N. Nikolaev}$^{2,3 \dag}$ and W.
Sch\"afer$^{2}$

\vskip 5mm

{\small

(1) {\it
Sobolev Institute of Mathematics,
    630090   Novosibirsk, Russia
}
\\

(2) {\it
Institut f. Kernphysik, Forschungszentrum J\"ulich, D-52425 J\"ulich, Germany
}
\\

(3) {\it
L.D.Landau Institute for Theoretical Physics, 142432 Chernogolovka, Russia
}
\\

$\dag$ {\it E-mail: N.Nikolaev@fz-juelich.de

}}

\end{center}

\vskip 5mm

\begin{abstract}

As Glauber has shown in 1959, the spin-flip phenomena caused by
the conventional spin-orbit interaction do vanish in the
scattering off heavy, strongly absorbing nuclei. On the other
hard,the origin of the s-channel helicity nonconservation (SCHNC)
in diffractive DIS is of the origin different from simple
spin-orbit interaction, and here we demonstrate that SCHNC in
vector meson production survives strong absorption effects in
nuclear targets. The intranuclear absorption often discussed in
terms of the saturation effects introduces a new large scale
$Q_A^2$ into the calculation of diffractive vector meson
production amplitudes. Based on the color dipole approach, we show
how the impact of the saturation scale $Q_A^2$ changes from the
coherent to incoherent/quasifree diffractive vector mesons.

\end{abstract}

\vskip 8mm

\section{Introduction: the fate of spin-orbit interaction for
heavy nuclei and the mechanism of SCHNC}

Heavy nuclei are strongly absorbing targets. Whether the spin-flip
effects in high energy scattering are washed out by this
absorption or not is not an obvious issue which we address in this
communication on an example of diffractive vector mesons.

The standard argument for vanishing of the spin-flip in elastic
scattering of spin ${1\over 2}$ particles off strongly absorbing
nuclei goes as follows: In the presence of the spin-orbit
interaction the scattering amplitude $f=A+2B\hat{\bs}\bn$, where
$\hat{\bs}$ is the spin operator, $\bn$ is the normal to the
scattering plane and the partial wave expansion of the
helicity-non-flip, $f_0$, and the helicity-flip, $f_1$, amplitudes
reads \cite{LandauLifshits} \bea f_0 &=& {1 \over 2ip} \sum_{l}
\{(l+1)[\exp(2i\delta_l^+)-1]
+l[\exp(2i\delta_l^-)-1]\}P_l(\cos\theta)\, ,\nonumber\\
f_1&=& {1 \over 2ip} \sum_{l}
[\exp(2i\delta_l^+)-\exp(2i\delta_l^-)]P_l^1(\cos\theta)\, ,
\label{eq:1.1} \eea where $\delta_l^{\pm}$ are the scattering
phases for $j=l\pm {1\over 2}$. In the presence of strong
absorption the scattering phases acquire large imaginary parts,
$\exp(2i\delta_l^{\pm}) \to 0$, and, consequently, for the
momentum transfer $\bDelta$ within the diffraction cone $f_1/f_0
\to 0$. More detailed treatment for elastic scattering of protons
off nuclei is found in Glauber's lectures \cite{Glauber}, the net
result is that the small contribution to the spin-flip amplitudes
comes only from the periphery of the nucleus, so that $f_1/f_0
\propto A^{-1/3}$, where $A$ is the mass number of a nucleus.

In high-energy QCD the diffractive production of vector mesons,
$$
\gamma^* p \to V p'\,, $$
proceeds via the exchange of colorless system of gluons in the $t$-channel.

\begin{figure}{h}

   \centering

   \epsfig{file=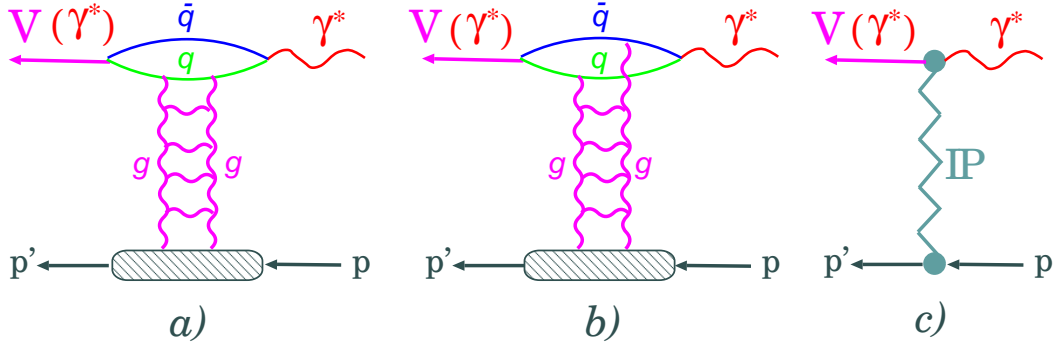,width=140mm}

   \caption{(a,b) The subset of two-gluon tower pQCD diagrams for
the pomeron exchange contribution (c) to the Compton scattering
(DIS) and diffractive vector meson production. Not shown are
two more diagrams with $q\leftrightarrow \bar{q}$.}

   \label{fig:Twogluontower}

\end{figure}
The fundamental property of such (multiple) gluon exchange is an
exact conservation of the s-channel helicity of quarks, recall the
similar property in the QED scattering of electrons in the Coulomb
field \cite{LandauLifshits2}. None the less,  QCD predicts a
non-vanishing helicity flip in diffractive production of vector
mesons off unpolarized nucleons \cite{IKspinflip,KNZspinflip}: the
origin of this SCHNC is in the subtle possibility that the sum of
helicities of the quark-antiquark pair in the diagrams of
fig.~(\ref{fig:Twogluontower}) can be unequal to the helicity of
photons and vector mesons. In the non-relativistic case the pure
S-wave deuteron with spin up consists of the spin-up proton and
neutron. However, in the relativistic case while the longitudinal
virtual photon contains the $q\bar{q}$ pair with
$\lambda_q+\lambda_{\bar{q}}=0$, the transverse photon with
helicity $\lambda_{\gamma}=\pm 1$ besides the $q\bar{q}$ state
with $\lambda_q+\lambda_{\bar{q}}=\lambda_{\gamma}=\pm 1$, also
contains the state with $\lambda_q+\lambda_{\bar{q}}=0$, in which
the helicity of the photon is carried by the orbital angular
momentum in the $q\bar{q}$ system. Furthermore, it is precisely
the  state $\lambda_q+\lambda_{\bar{q}}=0$ which gives the
dominant contribution to the absorption of transverse photons and
the proton SF $F_{2p}(x,Q^2)$ in the Bjorken limit. From the point
of view of the vector meson production, it is important that the
transverse and longitudinal $\gamma^*$ and $V$ share the
intermediate $q\bar{q}$ state with  $\lambda_q +
\lambda_{\bar{q}}=0$, which allows the s-channel helicity
non-conserving (SCHNC) transitions between the transverse
(longitudinal) $\gamma^*$ and longitudinal(transverse) vector
meson V. As a matter of fact, this mechanism of SCHNC does not
require an applicability of pQCD, B.G. Zakharov was the first to
introduce it in application to nucleon-nucleon scattering
\cite{SlavaSpinFlip}. The theoretical predicition of
energy-independent SCHNC in diffractive vector meson production
has been confirmed experimentally at HERA
\cite{ZEUSspinflip,H1spinflip}, the detailed comparison of the
theory and experiment is found in \cite{IgorPhD}.

Now notice that the argument about exact SCHC of quarks and
antiquarks applies equally to a one-pomeron exchange in the
scattering off a free nucleon and to multiple pomeron exchange in
the scattering off a nuclear target. Then for a sufficiently high
energy such that the lifetime of the $q\bar{q}$ fluctuation of the
photon, often referred to as the coherence time , and the
formation time of the vector meson are larger than the radius of
the nucleus $R_A$ (for instance, see
\cite{NNNcomments,BenharJPsi}), the above described origin of
SCHNC must be equally at work for the nuclear and free-nucleon
targets. In this communication we expand on this point, from the
practical point of view one speaks of the values of \beq
x={Q^2+m_V^2 \over 2\nu} < x_A \lsim 0.1\cdot A^{-1/3}\,.
\label{eq:1.2} \eeq Another property of interactions with nuclei
is the so-called saturation scale $Q_A$ which for partons with
$x\lsim x_A$ defines the transverse momentum below which their
density is lowered by parton fusion effects
\cite{MuellerSaturation,NonlinearKt,RajuReview}. It is interesting
to see how the emergence of the saturation scale affects the
$Q^2$-dependence of diffractive vector meson production, in
particular, its SCHNC properties. Here one must compare the
saturation scale $Q_A^2$ to the usual hard scale for diffractive
vector meson production \cite{NNZscan,IgorHardScale} \beq
\overline{Q}^2 \approx {1\over 4}(Q^2+m_V^2)\, . \label{eq:1.3}
\eeq

The further presentation is organized as follows. In section 2 we
start with the formulation of the color-dipole approach to
calculation of helicity amplitudes for diffractive vector meson
production. In section 3 we introduce the scanning radius and
comment on the potential sensitivity to the short-distance
behaviour of the wave function of vector mesons. In section 4 we
start a discussion of nuclear effects on an example of coherent
diffraction $\gamma^*A \to V A$ when the recoil nucleus remains in
the ground state. We demonstrate how the dependence $\propto
\overline{Q}^{-4}$ which is common to all helicity amplituides
changes to $\propto \overline{Q}^{-2}Q_A^{-2}$ for
$\overline{Q}^{2}\lsim Q_A^{2}$. Even stronger impact of
saturation is found for incoherent diffractive vector mesons
considered in section 5. In the Conclusions we summarize our
principal findings and comment on the possibilities of the COMPASS
experiment at CERN.

\section{The free nucleon target}

For the purposes of our discussion it is convenient to resort to
the color dipole formalism: the production process depicted in
fig.~(\ref{fig:Twogluontower}) factorizes into splitting of the
photon into $q\bar{q}$ dipole way upstream the target, s-channel
helicity conserving elastic scattering of the dipole off a target,
and projection of the $q\bar{q}$ dipole onto the vector meson
state. We restrict ourselves to the contribution from the
$q\bar{q}$ Fock states of the vector meson which is a good
approximation for $x\sim x_A$. The momentum-space calculation of
the helicity amplitudes has been worked out time ago in
\cite{INSDwave,IgorPhD}, the crucial ingredient in preserving the
rotation invariance is the concept of the running polarization
vector for the longitudinal vector mesons. Following \cite{NZ91},
we make the Fourier transform to the color-dipole space and
represent the helicity amplitudes ${\cal A}_{fi}(x,\bDelta)$,
where $i=\lambda_{\gamma}$ and $f=\lambda_V$ are helicities of the
initial state photon and the final state vector meson,
respectively, in the color dipole factorization form \beq {\cal
A}_{fi}(x,\bDelta)= \langle V_f |{\cal A}_{q\bar{q}}(\br,\bDelta)
|\gamma^*_i\rangle = i\int_0^1 dz \int d^2\br \sigma(\br,\bDelta)
\exp[{i\over 2}(1-2z)(\br \bDelta)] I_{fi}(z,\br)\, ,
\label{eq:2.1} \eeq where $\bDelta$ is the transverse momentum
transfer in the $\gamma^* \to V$ transition, $I_{fi}(z,\br)=
\Psi^*_{V,f}(z,\br)\Psi_{\gamma^*,i}(z,\br)$ and the summation
over the helicities $\lambda,\overline{\lambda}$ of the
intermediate $q\bar{q}$ pair is understood. The wave function
$\Psi_{V,f}(z,\br)$ of the final state vector meson contains the
spin-orbital part and the "radial" wave functions, defined in
terms of the vertex function $\Gamma_V(z,\bk)\bar{q} S_{\mu}q
V_{\mu}$, where $S_{\mu}$ is the relevant Dirac structure and
$V_{\mu}$ is the running polarization vector which must be so
defined as to guarantee the rotational invariance for the fixed
invariant mass $M$ of the lightcone $q\bar{q}$ Fock state of the
vector meson \cite{INSDwave,IgorPhD},
$$
M^2 = {m_f^2 +\bk^2 \over z(1-z)}\, ,
$$
where $\bk$ is the transverse momentum of the quark in the vector
meson and $z$ and $(1-z)$ are fractions of the lightcone momentum
of the vector meson carried by the quark and antiquark,
respectively. In the momentum-space $\psi_V(z,\bk) \propto
\Gamma_V(z,\bk)/(M^2-m_V^2)$, the Fourier transform to the dipole
space depends on the Dirac structure $S_{\mu}$. For the sake of
simplicity, here we take $S_{\mu}=\gamma_{\mu}$, the exact form
for the pure $S$ and $D$ wave states is found in
\cite{INSDwave,IgorPhD}, the major conclusions on the impact of
nuclear absorption do not depend on the exact form of $S_{\mu}$.
If one defines the radial wave functions for the transverse (T)
and longitudinal (L) vector mesons as \beq \psi_T(z,\br)=\int
d^2\bk \psi(z,\bk)\exp(i\bk\br)\, ,~~~~ \psi_L(z,\br)=\int d^2\bk
M \psi(z,\bk)\exp(i\bk\br)\, , \label{eq:2.2} \eeq then \bea
I_{LL} &=&  4Qz^2(1-z)^2 K_0(\varepsilon r)\psi_L(z,r)\\
\label{eq:2.3}
I_{TT} & =& m_f^2 K_0(\varepsilon r)\psi_T(z,r) - [z^2+(1-z)^2]
 \varepsilon K_1(\varepsilon r)\psi_T'(z,r)\\
\label{eq:2.4}
I_{LT}& =&-i2z(1-z)(1-2z)\psi_L(z,r)\varepsilon K_1(\varepsilon r){(\bfe \br)\over r}\\
\label{eq:2.5}
I_{TL} &=& -i 2Qz(1-z)(1-2z) K_0(\varepsilon r)\psi_T'(z,r){(\bV^* \br)\over r} \\
\label{eq:2.6} I_{TT'}&=&4z(1-z) \varepsilon K_1(\varepsilon
r)\psi_T'(z,r){(\bfe \br)^2\over r^2} \label{eq:2.7} \eea where
$\psi_T'(z,r) = \partial \psi_T(z,r)/\partial r$, the polarization
vectors $\bfe$ and $\bV$ are for the transverse photon and vector
meson, respectively, $\varepsilon^2 = z(1-z)Q^2+m_f^2$, the Bessel
functions $K_{0}(x)$ and $K_{1} = K_0'(x)$ describe the lightcone
wave function of the photon, $I_{TT}$ and $I_{TT'}$ describe the
helicity-non-flip and double-helicity-flip production of
transverse vector mesons by transverse photons, in the latter case
$\bV^* = \bfe$.

The off-forward generalization of the color-dipole scattering
amplitude has been introduced in \cite{NNPZZslopeVM} \bea
&&\sigma(\br,\bDelta)= {2\pi \over 3}\int {d^2\bkappa \over
(\bkappa -{1\over 2}\bDelta)^2  (\bkappa +{1\over 2}\bDelta)^2}
{\cal F}(x,\bkappa,\bDelta) \alpha_S(\bkappa^2)
\nonumber\\
&&\times  \left\{[1-\exp(i\bkappa\br)]\cdot[1 -\exp(-i\bkappa\br)]
-[1-\exp({1\over 2}i\bDelta\br)]\cdot[1-\exp(-{1\over
2}i\bDelta\br)] \right\} \label{eq:2.8} \eea Here ${\cal
F}(x,\bkappa,\bDelta)$ is the off-forward unintegrated
differential gluon structure function of the nucleon, the gross
features of its $\bDelta$-dependence are discussed in
\cite{NNPZZslopeVM}.

The analysis of \cite{NNPZZslopeVM} focused on the $LL$ and $TT$
amplitudes, in which case the dominant contribution comes from
$z\sim {1\over 2}$ and corrections to the $\bDelta$-dependence
from the factor $\exp[{i\over 2}(1-2z)(\br \bDelta)]$ in
(\ref{eq:2.1}) can be neglected. This factor is crucial, though,
for the helicity-flip transitions. For small dipoles and within
the diffraction cone the leading components of the  $LT$ and $TL$
amplitudes come from the second term in the expansion \beq
\exp\Big({1\over 2}i (1-2z) (\bDelta \br)\Big) = 1 +{1\over 2}i
(1-2z)(\bDelta \br) \label{eq:2.9} \eeq so that upon the azimuthal
averaging the effective integrands take the form \bea
I_{LT} &=& {1\over 2} z(1-z)(1-2z)^2\psi_L(z,r)\varepsilon r K_1(\varepsilon r)(\bfe \bDelta)\\
\label{eq:2.10} I_{TL} &=& {1\over 2}Qz(1-z)(1-2z)^2
K_0(\varepsilon r) r \psi_T'(z,r)(\bV^* \bDelta) \label{eq:2.11}
\eea In the case of the double-flip $TT'$ amplitude $\bfe^2 =0$
and one needs to expand the integrand up to the terms $\propto
(\br \bDelta)^2$: \beq \exp\Big({1\over 2}i (1-2z)(\bDelta
\br)\Big){(\bfe \br)^2 \over r^2} \sigma(x,\br,\bDelta)
\Rightarrow {1\over 60}(\bfe\bDelta)^2
r^2[\sigma(\infty,0)+(1-2z)^2\sigma(r,0)] \label{eq:2.12} \eeq so
that the corresponding integrand of the double-flip amplitude will
be of the form \bea I_{TT'} ={1\over 15}z(1-z) \varepsilon r^2
K_1(\varepsilon r) \psi_T'(z,r)(\bfe
\bDelta)^2[\sigma(\infty,0)+(1-2z)^2\sigma(r,0)] \label{eq:2.13}
\eea

\section{The scanning radius, hard scale expansion and sensitivity to the
short distance wave function of the vector meson}

%%%%%%%%%%     section 3

Consider the pQCD regime of large $Q^2$. The useful representation
for small color dipoles is \cite{NZglue} \bea \sigma(\br,0)& =&
{\pi^2 \over 3}r^2 \alpha_S(q^2) G(x,q^2),~~~ q^2\approx {10\over
r^2}\, . \label{eq:3.1} \eea Then, because of the exponential
decrease, $K_{0,1}(\varepsilon r) \propto \exp(-\varepsilon r)$,
the amplitudes for the free nucleon target will be dominated the
contribution from $r=r_S$, where the scanning radius \beq r_S \sim
{a_S\over \varepsilon} \approx {a_S\over \overline{Q}} = {2 a_S
\over \sqrt{Q^2 + m_V^2}}\,, ~~~ a_S \approx 3\, . \label{eq:3.2}
\eeq

The simplest case is that of the $LL$ amplitude:
\bea
{\cal A}_{LL} &\propto& Q r_S^2 \sigma(r_S,0)  K_{0}(a_S)
\int dz z^2(1-z)^2 \psi_L(z,r_S) \nonumber\\
&\propto& {Q \over \overline{Q}^4} \alpha_S(\overline{Q}^2)
G(x,\overline{Q}^2) \propto  {Q G(x,\overline{Q}^2)\alpha_S(\overline{Q}^2)
 \over (Q^2+m_V^2)^2}\,.
\label{eq:3.3}
\eea
The expansion in powers of the scanning radius $r_S$ is an expansion in
inverse powers of the hard scale $\overline{Q}$.
Here one factor $1/(Q^2+m_V^2)$ is the same as in the Vector Dominance Model,
in the color dipole language it can be identified with the overlap of the
photon and vector meson wave functions, the second factor  $1/(Q^2+m_V^2)$
derives from the pQCD form (\ref{eq:3.1}) of the dipole cross section.

The helicity-flip amplitudes will be of the form
\bea
{\cal A}_{LT} &\propto&
(\bfe \bDelta) r_S^2 \sigma(r_S,0) a_S K_1(a_S)
\int dz z(1-z)(1-2z)^2\psi_L(z,r_S)\\
\label{eq:3.4} {\cal A}_{TL} & \propto&(\bV^* \bDelta) Q  r_S^3
\sigma(r_S,0) K_0(a_S)\int dz z(1-z)(1-2z)^2 \psi_T'(z,r_S)
\label{eq:3.5} \eea Notice the senisitivity to the short-distance
behavior of the vector meson wave wave function in the last
result. The soft, oscillator-like interaction would give the wave
function which at short distances is a smooth function of $\br^2$,
to that \beq \psi_T'(z,r) \sim -{r\over R_V^2} \psi_T(z,0)
\label{eq:3.6} \eeq whereas the attractive Coulomb inetraction at
short distances suggests \cite{NNZscan,NNPZdipoleVM,NNPZcdVMsyst}
"hard", Coulomb-like $\psi_T(z,r) \propto \exp(-r/R_C)$  when \beq
\psi_T'(z,r) \sim -{1\over R_C} \psi_T(z,0) \label{eq:3.7} \eeq In
the case of the soft short-distance wave function  the
helicity-flip amplitude ${\cal A}_{TL}$ would acquire extra small
factor $r_S/R_V \propto 1/(R_V\overline{Q})$. The similar
discussion is relevant to the contribution to the ${\cal A}_{TT}$
from the term $ \propto - \varepsilon K_1(\varepsilon r)
\psi_T'(z,r)$ in $I_{TT}$, eq. (\ref{eq:2.4}), which for the hard
short-distance wave function is larger and somewhat enhances the
transverse cross section $\sigma_T$ and lowers
$\sigma_L/\sigma_T$, see also the discussion in
\cite{KolyaCracow}. Finally, the leading term of expansion in
powers of $r_S$ of the double-helicity-flip amplitude is of the
form \beq {\cal A}_{TT'} \propto (\bfe \bDelta)^2 z (1-z) r_S^3
a_S  K_1(a_S)\sigma(\infty,0) \int dz z (1-z)\psi'_T(z,r_S)\, .
\label{eq:3.8} \eeq It is proportional to the dipole cross section
for the nonperturbative large color dipole \cite{KNZspinflip}

\section{Nuclear saturation effects in coherent diffraction}

%%%%%%%%%%%   section 4

In the coherent diffractive production of vector mesons the target
nucleus remains in the ground state,
$$
\gamma^* A \to VA\,
$$
For heavy nuclei such that their radii are much larger than the
dipole size and the diffraction slope in the dipole-nucleon
scattering, only the forward dipole-nucleon scattering enters the
calculation of the nuclear profile function. Compared to the
conventional derivation of the Glauber formulas for the nuclear
profile functions, there are little subtleties with the presence
of the phase factor $\exp[{i\over 2}(1-2z)(\br \bDelta)]$ in the
color dipole factorization formula (\ref{eq:2.1}), but a careful
rederivation gives the nuclear diffractive amplitudes of a form
\beq {\cal A}_{fi}=2i\int d^2\bb \,\exp(-i\bDelta \bb) \langle
V_f|\left\{1-\exp[-{1\over
2}\sigma(r,0)T(\bb)]\right\}|\gamma^*_i\rangle \exp[{i\over
2}(1-2z)(\br \bDelta)]\,, \label{eq:4.1} \eeq where $T(\bb)=\int
dz n_A(\bb,z)$ is the standard nuclear optical thickness at an
impact parameter $\bb$. A comparison with the free nucleon
amplitude (\ref{eq:2.1}) shows that the nuclear profile function
\beq \Gamma(\bb,\br)= 1-\exp[-{1\over 2}\sigma(r,0)T(\bb)]
\label{eq:4.2} \eeq plays the r\^ole of the color dipole cross
section per unit area in the impact parameter space.

First of all, we notice that the color dipole dependence of the
ovelap of wave functions of the photon and vector meson does not
change from the free nucleon to the nuclear case. The same is true
of the calculation of expectation values over the orientation of
color dipoles, see eqs. (\ref{eq:2.10} - \ref{eq:2.13}). The $r$
dependence of the integrands changes, though.

Ref. \cite{NonlinearKt} gives the detailed discussion of the
reinterpretation of the nuclear profile function in terms of the
saturating nuclear gluon density
\cite{MuellerSaturation,RajuReview} and of the limitations of such
an interpretation for observables more complex than the single
particle spectra. For the purposes of our discussion it is
sufficient to know that in terms of the so-called saturation scale
\beq Q_A^2(\bb) = {4\pi^2 \over 3}\alpha_S(Q_A^2)G(Q_A^2)T(\bb)
\label{eq:4.3} \eeq the nuclear attenuation factor in
(\ref{eq:4.2}) can be represented as \beq \exp[-{1\over
2}\sigma(r,0)T(\bb)] \approx \exp[-{1\over 8}Q_A^2 r^2]
\label{eq:4.4} \eeq

Let the nucleus be very heavy such that the saturation scale
$Q_A^2$ is very large (for the estimates of $Q_A^2$ for realistic
nuclei see \cite{NonlinearKt}). For color dipoles with $r^2 >
r_A^2 ={8/Q_A^2}$ the nucleus is opaque, i.e., we have a saturated
$\Gamma(\bb,\br) \approx 1$ independent on the dipole size. The
new large scale $Q_A^2$ must be compared to $\overline{Q}^2$ of
eq. (\ref{eq:1.3}).

First, there is a trivial case of $\overline{Q}^2 \gg Q_A^2$.
In this case $r_S^2 \ll r_A^2$, so that $2\Gamma(\bb,\br)=
\sigma(r_s,0)T(\bb)$, i.e., the impulse approximation is at work
and the nuclear amplitude has precisely the
same structure as the free nucleon one,
\beq
{\cal A}^{(A)}_{fi}(\bDelta) = {\cal A}^{(N)}_{fi}(\bDelta)
\int d^2\bb \exp(-i\bb\bDelta)T(\bb) =
{\cal A}^{(N)}_{fi}(\bDelta)\cdot A\cdot G_{em}(\bDelta)\,,
\label{eq:4.5}
\eeq
where $G_{em}(\bDelta)$ is the charge form factor of a
nucleus.

Much more interesting is the case of coherent diffractive DIS at
$Q^2 \ll Q_A^2$, when the color dipoles in the photon have the
dipole size $r \sim {1\over Q} \gg r_A^2$, i.e., DIS proceeds in
the regime of opacity and saturated color dipole cross section per
unit area in the impact parameter space. Apart from this
difference all the arguments of section 3 on the scanning radius
will be fully applicable, only the scale for the scanning radius
will change to $a_S \approx 1$. However, the, the power expansion
in $r_S$ will change: instead of the common factor $r_S^4$ of
section 3 for the free nucleon target one obtains $r_S^2 r_A^2$
for diffraction off nuclei in the saturation scale, i.e., the
substitution \beq \left.{1\over \overline{Q}^4}\right|_N
\Rightarrow \left.{1\over \overline{Q}^2 Q_A^2}\right|_A \,
\label{eq:4.6} \eeq in the common prefactor of all helicity
amplitudes. Otherwise there is no nuclear mass number dependent
suppression of the relative strength of the helicity-flip and
non-flip amplitudes compared to the free nucleon case. Finally,the
$\bDelta$ dependence of the effective nuclear form factor will be
close to the  $\bDelta$ dependence of an amplitude of elastic
scattering on a black disc, \beq G_{em}(\bDelta) \Rightarrow
{J_1(R_A\Delta) \over R_A\Delta} \label{eq:4.7} \eeq

We emphasize that although the presence of those nuclear
form factors limit the practical observation of  coherent
diffractive DIS to momentum transfers within the nuclear
diffraction cone, $\bDelta^2 \lsim R_A^{-2}$, and these
small $\bDelta$ cause the kinematical suppression of the
helicity-flip within the coherent cone, there is no
nuclear suppression of helicity flip even on a black nucleus.

\section{Incoherent/quasielastic diffraction}

%%%%%%%%%    section 5

In the incoherent (quasielastic, quasifree) diffractive vector
meson production
$$
\gamma^* A \to V A^*
$$
one sums over all excitations and breakup of the target nucleus
without production of secondary particles. The process looks like
a production off a quasifree nucleon of the target subject to
certain intranuclear distortions of the propagating dipoles. The
relevant multichannel formalism has been worked out in
\cite{KolyaJETP}, the generalization to the color dipole formalism
for $z\approx {1\over 2}$ is found in
\cite{NNNcomments,BenharJPsi}. Here we notice that in the color
dipole language, the calculation of the helicity amplitudes will
be exactly the same as for the free nucleon target but with the
extra attenuation factor of eq. (\ref{eq:4.4}) in all the
integrands. i.e., the inoherent differential cross section  equals
\beq {d\sigma(\gamma^*_i A \to V_f A^*)\over d\bDelta^2} =\int
d^2\bb T(\bb) {d\sigma_{qel}\over d\bDelta^2}\, , \label{eq:5.1}
\eeq where the amplitudes of the quasielastic production off a
quasifree nucleon are given by
 \beq {\cal
A}_{fi}^{(qel)}(x,\bDelta)= i\int_0^1 dz \int d^2\br
\sigma(\br,\bDelta)\exp[-{1\over 2}\sigma(r,0)T(\bb)] \exp[{i\over
2}(1-2z)(\br \bDelta)] I_{fi}(z,\br)\, , \label{eq:5.2} \eeq

In the genuine hard regime of $Q^2 \gg Q_A^2$ the nuclear
attenuation can be neglected and one recovers the free nucleon
cross section times the number of nucleons. In the opposite regime
of strong saturation, $\overline{Q}^2 \ll Q_A^2$, the $r$
dependence of the attenuation factor is stronger than that of the
photon wave functions $K_{0,1}(\varepsilon r)$. Then, repeating
the derivation of the scanning radius in section 3, one will find
\beq r_S^2 \approx {3\over 2} r_A^2\, . \label{eq:5.2*} \eeq

The functional dependence of helicity amplitudes on the scanning
radius $r_S$ will be the same as for the free nucleon target with
one exception. Namely, the Bessel functions in the photon wave
function shall enter with the argument \beq a_S = \varepsilon r_S
\approx {\sqrt{3}\, \overline{Q} \over Q_A} \ll 1. \label{eq:5.3}
\eeq In this limit \beq K_0(a_S) \approx \log\Big({Q_A \over
\overline{Q}}\Big) \label{eq:5.4} \eeq which shows that some of
the amplitudes will have a logarithmic enhancement, whereas \beq
K_1(a_S) \approx {Q_A \over \overline{Q}} \label{eq:5.5} \eeq is
indicative of even stronger enhancement. However, the closer
inspection of the helicity-flip amplitude ${\cal A}_{LT}$
(\ref{eq:2.10}) shows that $K_1(a_S)$ enters as a product $a_S
K_1(a_S)$ which is a smooth function at $a_S \ll 1$. The
helicity-flip amplitude ${\cal A}_{TL}$ of eq. (\ref{eq:2.11})
exhibits only the weak logarithmic enhancement (\ref{eq:5.4}). The
case of the helicity-non-flip ${\cal A}_{TT}$ and double-flip
${\cal A}_{TT'}$ is a bit more subtle. Here one encounters \beq -
\varepsilon K_1(\varepsilon r_S)\psi_T'(z,r_S) \sim -{1\over r_S}
 \psi_T'(z,r_S)
\label{eq:5.6} \eeq which for the soft short-distance wave
function can be estimated as \beq {1\over R_V^2}\psi_T(z,r_S)
\label{eq:5.7} \eeq whereas for the hard Coulomb wave function one
finds an enhancement \beq {Q_A \over R_C} \psi_T(z,r_S)
\label{eq:5.8} \eeq

To summarize, strong nuclear absorption does not generate any
special suppression of the helicity-flip amplitudes compared to
the non-flip ones. Furthermore, the estimate (\ref{eq:5.8})
suggests even a possibility of an enhancement of the double-flip
transitions depending on the hardenss of the short-distance wave
function of the vector meson. Finally, this discussion shows that
in the saturation regime  the saturation scale $Q_A^2$ becomes the
hard factorization scale for incoherent diffractive production.
Namely, for $\overline{Q^2} \ll Q_A^2$ this amounts to the
substitution \beq \left.{1\over \overline{Q}^4}\right|_N
\Rightarrow \left.{1\over Q_A^4}\right|_A \, \label{eq:5.9} \eeq
in the common prefactor of all helicity amplitudes. Similarly, the
diffraction slope for the vector meson production will be the same
as that  for the free nucleon target but taken for the hard scale
$Q_A$.

Here we focused on the single incoherent scattering approximation.
The higher order incoherent interactions can readily be treated
following the technique of ref. \cite{KolyaJETP}, they wouldn't
change major conclusions on the interplay of the DIS hard scales
$\overline{Q}^2$ and the saturation scale $Q_A^2$.

\section{Conclusions}

Our principal finding is a lack of nuclear suppression
of the helicity-flip phenomena in
hard diffractive production off strongly absorbing nuclei,
which is in striking contrast to the familiar strong
nuclear attenuation of the spin-orbit interaction effects
as predicted by the Glauber theory.
The QCD mechanism behind this finding is that
absorption only affects the color dipole-nucleus scattering
amplitude in which the $s$-channel helicity of the quark
and antiquark is anyway conserved exactly. The helicity
flip originates from the relativistic mismatch of the
sum of helicities of the quark and antiquark and the
helicity of the vector meson and photon. Within the color
dipole approach we demonstrated how the expansion of
helicity amplitudes in powers of the scanning radius
and saturation radius changes from the free nucleon to
coherent nuclear to incoherent (quasielastic) nuclear
diffractive production.

It is proper to recall the early claims by Greenberg and Miller of
the so-called vector color transparency - the vanishing spin-flip
phenomena in hard processes on nuclear targets \cite{VectorCT}.
These authors considered the polarization of ejected protons in
quasielastic scattering $A(e,e'p)$ off nuclei in the version of
spin-orbit interaction model and overlooked Zakharov's helicity
mismatch mechanism which, as we demonstrated above, persists in
hard scattering. We discussed diffractive DIS, but all the
arguments are equally applicable to the hadron-nucleus scattering.

The coherent and incoherent diffractive vector meson
production off nuclei can be studied experimentally in
the COMPASS experiment at CERN \cite{COMPASS}. Whereas
we are confident in our predictions for perturbatively large
saturation
scale $Q_A^2$, the numerical estimates for the saturation
scale give disappointingly moderate $Q_A^2 \sim 1$ GeV$^2$.
None the less, the  qualitative pattern of predicted changes of
the $Q^2$ dependence of diffractive vector meson production
from the free nucleon to coherent nuclear and incoherent
nuclear cases must persist even at moderately large $Q_A^2$.

This work was partly supported by the grant INTAS 00366.

%%%%%%%%%%%%%%%%%%%%%%%%%%%%%%%%%%%%%%%%%%%%%%%%%%

%\begin{wrapfigure}{R}{8cm}

%\mbox{\epsfig{figure=fig1.eps,width=7.8cm,height=6cm}}

%{\small{\bf Figure 1.} Figure title aaaaaaa aaa aaa aaa aaa aaaaaa aaaaaaaaaaa aaa}

%\medskip

%\end{wrapfigure}

%To insert figure (with the help of wrapfig.sty)\\

%%%%%%%%%%%%%%%%%%%%%%%%%%%%%

\end{document}